\documentclass[letterpaper,twocolumn]{article}
\usepackage{geometry}
\usepackage[raggedright]{titlesec}
\usepackage{cite}
\usepackage{hyperref}
\usepackage[flushleft]{threeparttable}
\usepackage{epstopdf}
\usepackage{changepage}

\usepackage{microtype}
\usepackage[utf8]{inputenc}
\newcommand\MyHead[2]{%
  \multicolumn{1}{|l|}{\parbox{#1}{\centering #2}}}
\usepackage{multirow}
\usepackage[labelfont=bf,labelsep=period,justification=raggedright]{caption}
\usepackage{amsmath}
\usepackage{graphicx}
\usepackage{url}

\date{}

\titlespacing{\section}{0pt}{*2.0}{*2.0}
\titlespacing{\subsection}{0pt}{*1.8}{*1.8}

\begin{document}

\title{\Large\textbf{A pipeline for identifying integration sites of mobile elements in the genome using next-generation sequencing}\normalsize}

\author {Raunaq Malhotra$^{1,\ast}$, Daniel Elleder$^{4}$, Le Bao$^{2}$, David R Hunter$^{2}$, Mary Poss$^{3}$, and Raj Acharya$^{1}$\\
\bf $^1$Department of Computer Science and Engineering, $^2$Department of Statistics, \\ 
\textbf{$^3$Department of Biology,}
\textbf{Pennsylvania State University}, University Park, PA, 16802, USA\\
\textbf{$^4$Institute of Molecular Genetics, Academy of Sciences of the Czech Republic,}\\ Prague, 14220, Czech Republic
}
\maketitle 

\thispagestyle{empty}

\begin{center}
\large\textbf{Abstract}
\end{center}

\vspace{-2mm}
Next Generation Sequencing (NGS) reads obtained by sequencing of the junction of a mobile element and the host flanking region from individuals in a population are typically mapped to a reference genome to determine the location of the mobile element-host junction. We propose a clustering pipeline for grouping such NGS data into clusters corresponding to the locations of integration sites in the genome. Our pipeline relies on the UCLUST clustering software, which clusters reads into groups using a clustering threshold, to cluster the integration sites NGS reads into groups based on their site of origin. An optimal clustering threshold is chosen based on a proposed clustering measure, $I-index$. We evaluate our pipeline on simulated integration sites data from the human genome and compare its performance to UCLUST clustering. Our pipeline is more accurate in recovering both the number and the correct sequence of the integration sites when compared to the other method. This pipeline can be beneficial in detecting the mobile element-host junctions in a population for species with no reference genome.

\medskip
\noindent
\vspace{-20pt}
\section{Introduction} 
The genomes of mammals contain several types of repetitive elements capable of mobilizing to new locations (known as mobile elements). This process is known as transposition and it leads to differences in the distribution of specific classes of mobile elements in the genome of individuals of a species, and as has been recently documented, within cells of the same individual \cite{Ishida}. Because mobile elements can affect the function and structure of the genome region near the site of integration, they have been linked to disease in both humans and mice \cite{belancio2008mammalian}. There is therefore, considerable interest in identifying the location of these elements in the host genome. 

These studies can be performed by amplifying a small segment of DNA that spans a terminus of the mobile element genome and the adjacent host genome region \cite{bao2014computational}. NGS technologies provide a high throughput approach to comprehensively investigate the differences in location of specific mobile element among individuals. The output from the NGS platform is a large collection of reads that provide high sequence coverage of the short junction fragments for each element in the genome, where the reads from different individuals can be bar-tagged to distinguish them into individuals. 

\begin{figure}[!htb]
\centering
\includegraphics[scale=0.25]{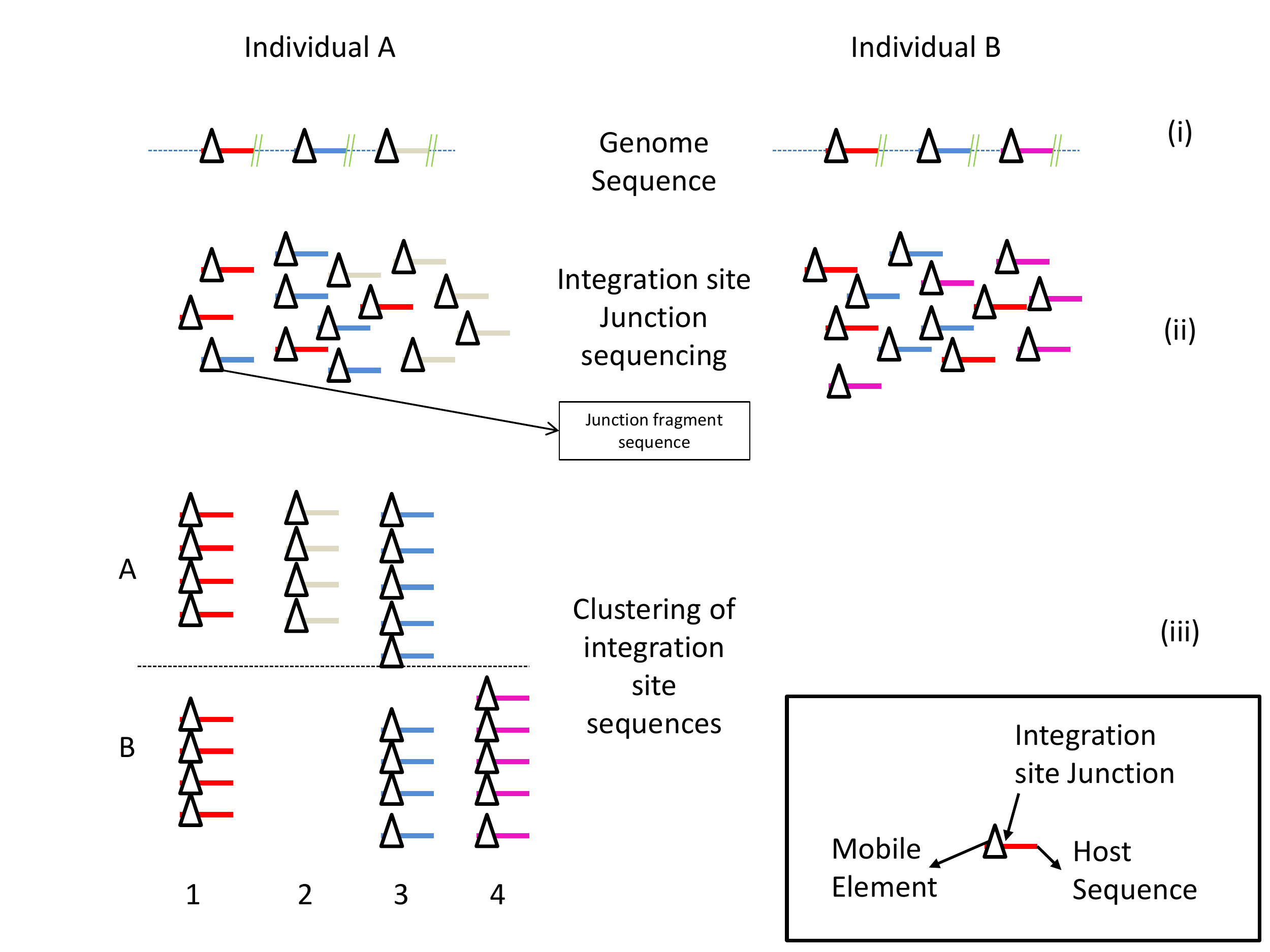}
\caption{\label{fig:01}  A pictorial example of the clustering pipeline for integration site NGS data from two individuals.  }
\end{figure}

For organisms with a high quality reference genome (eg. human and mouse), the NGS reads can be mapped to a reference sequence to identify the location of integration sites in the genome\cite{li2012mouse,giordano2007new}. An alternative approach for analyzing such junction fragment data is to cluster the reads (Figure \ref{fig:01}). The number of clusters determines the number of integration sites sequenced and the consensus sequence of the reads from a cluster represents the sequence of the integration site (inset in Figure \ref{fig:01}). Although the location of integration of the integration site is not known via clustering methods, these methods have advantage over mapping methods, as they do not require a reference genome and are more tolerant to sequence differences amongst multiple individuals. Moreover, clustering methods can be used for non-model organisms where a reference genome is not available, or in cases where assembly is not complete. 

A clustering algorithm groups reads that have a similarity score derived from pairwise alignment exceeding a certain threshold. The choice of the threshold is important as a high value can split reads from the same integration sites into multiple clusters thereby increasing the number of integration sites reported, while a low clustering threshold may combine reads from two or more integration sites leading to under reporting of the integration sites. It is important that the clusters obtained are compact in that reads in a cluster are all similar to each other, and also that the clusters are well separated from each other. This can be evaluated using clustering compactness measures \cite{dunn,dbindex}. The goal of this study is to develop an approach to improve the accuracy of clustering methods used to identify integration sites of mobile elements.

We present a pipeline that employs the UCLUST algorithm in USEARCH \cite{usearch} for clustering reads from an integration site dataset and optimizes the clustering threshold using an internal cluster compactness measure $(I-index)$. The $I-index$ is derived from two cluster compactness measures (Dunn Index and Davies-Bouldin Index) \cite{dunn,dbindex}. The pipeline is evaluated on two simulated datasets that differ in sequencing depth and the results compared to those obtained from the UCLUST algorithm alone using cluster external and internal compactness measures. The results indicate that the proposed pipeline has improved compactness measures, clusters reads into a collection of integration sites, and outperforms a single round of clustering at any clustering threshold. Sequence depth does not affect clustering performance but the sequence content of a dataset, specifically sequences from repeat regions, impacts the overall clustering accuracy. 

\vspace{-10pt}
\section{Methods}
\subsection{Definitions}
We denote the collection of reads obtained from an integration site NGS sequencing project containing $P$ different integration site junctions as $\mathbf{R}=\{R_1,R_2,...R_N\}$. A clustering algorithm $\mathit{C} (\theta):\mathbf{R} \rightarrow \{ M_1,M_2,\ldots, M_K\}$ defines a map from the reads $\mathbf{R}$ to the $K$ clusters, wherein each read $R_i$ is assigned to a cluster $\mathit{C} (\theta,R_i)$ based on the clustering threshold $\theta$. A consensus sequence of reads assigned to a cluster $M_i$, denoted as $Con(M_i)$, is computed by majority base calling from a position by position alignment of the reads assigned to a cluster.



\subsection{Clustering Pipeline}
The number of integration sites ($P$) is unknown, and the goal of the clustering pipeline is to obtain a collection of $K \approx P$ clusters, with their consensus sequence $Con(M_i)$ being identical to the sequence of the integration site. To achieve this goal, reads sequenced from a single integration site should be grouped into a single cluster. As such reads only differ either by sequencing errors or polymorphisms in the host, they are amenable to global alignment and clustering algorithms can be used for their grouping. 
 
We will refer to an invocation of $\mathit{C}(\theta)$ as a clustering run. The pipeline consists of four major steps. In the first step, we cluster the collection of reads $\mathbf{R}$ at clustering threshold $\theta^{(1)}$ using the clustering algorithm $\mathit{C} (\theta^{(1)})$. The clustering algorithm generates a collection of clusters $\mathcal{M}^{(1)}(\theta^{(1)}) = \{ M_1^{(1)},M_2^{(1)},\ldots,M_K^{(1)} \}$ (The superscript 1 denotes that the clusters are obtained in step one of the pipeline). We also compute the consensus sequences $ Con(\mathcal{M}^{(1)})$ for each cluster in $\mathcal{M}^{(1)}$ based on the reads assigned to it. In the second step, we obtain a subset of clusters $\mathcal{M}^{(2)} \subset \mathcal{M}^{(1)}$ by only selecting clusters in $\mathcal{M}^{(1)}$ containing greater than $T_{err}$ number of reads, namely, $\mathcal{M}^{(2)}(\theta^{(1)},T_{err}) = \{M_i^{(1)}, |M_i^{(1)}| >T_{err}\}$. In step three, we re-cluster the consensus sequences of $\mathcal{M}^{(2)}$ clusters $Con(\mathcal{M}^{(2)})$ using the clustering algorithm $\mathit{C}(\theta^{(3)})$ to obtain a revised cluster set $\mathcal{M}^{(3)}(\theta^{(3)}) = \{M_1^{(3)},M_2^{(3)},\ldots,M_{K'}^{(3)} \}$. Next we map all the reads in $\mathbf{R}$ to the consensus sequences of $\mathcal{M}^{(3)}$ obtained from step three. The clusters are updated to the consensus sequences of their constituent reads. The cluster set $\mathcal{M}^{(3)}$ obtained in step four constitutes the final set of clusters obtained from read set $\mathbf{R}$ using the pipeline. The consensus sequences of the clusters $\mathcal{M}^{(3)}$ represents the sequences of the integration sites constituting the collection of reads $\mathbf{R}$ . The reads assigned to a cluster in  $\mathcal{M}^{(3)}$ denote the reads sampled from the integration site represented by the cluster's consensus sequence. 


The rationale for removing small clusters (with reads less than $T_{err}$) in step two is that they likely correspond to splitting of reads from an integration site junction into two or more clusters in $\mathcal{M}^{(2)}$. These small clusters arise due to the presence of sequencing errors in the reads and at the clustering threshold $\theta^{(1)}$, reads from the same integration site are split into more than one cluster leading to a false increase in the apparent number of integration sites present in the dataset. The rationale for clustering in step three is to combine different clusters that originated due to sequencing errors, but actually correspond to a single integration site. The clustering threshold in step three $\theta^{(3)}$ can be different from the one used in step one. Step four reassigns every single read to the clusters obtained in step three. This step ensures that the clusters and reads removed in step two are also assigned to the integration site from which they were sampled. 

The clustering algorithm $\mathit{C}(\theta)$ can be any available clustering tool \cite{usearch,cd-hit}. We use the clustering algorithm UCLUST \cite{usearch} based on its favorable comparisons to other clustering softwares on speed and relative accuracy \cite{usearch}. UCLUST takes the reads and a clustering threshold ($\theta$) for clustering reads, which is what is required for our clustering pipeline. The computational complexity for UCLUST is linear in the number of the reads, which is also beneficial for fast clustering. The options for UCLUST algorithm in USEARCH were set to: global matching, nofastalign, maxrejects, maxaccepts as 0, identity definition to 1. For mapping of reads in step 4, the query mapping to database option in USEARCH was used with parameters set to: global matching, nofastalign, maxrejects and maxaccepts to 80, identity definition 1.    

The expected computational complexity of a fast clustering algorithm $\mathit{C}(\theta)$ is linear in the number of reads. The computational complexity of computing alignment score is quadratic in the read length in the worst case. Thus a clustering run has computational complexity of $O(|\mathbf{R}|\cdot |R_{i}|^{2})$. Our pipeline performs two rounds of clustering, where in step 3, only $|\mathcal{M}^{(3)}|$ sequences are clustered. Thus, the overall complexity of the pipeline is of the order of $O((|\mathbf{R}|+|\mathcal{M}^{(3)}|) \cdot |R_i|^2 )$, which is also linear in the number of reads. 


\subsection{Determining the clustering thresholds}
There are three parameters in our pipeline, namely $(\theta^{(1)},T_{err},\theta^{(3)})$ that determine the collection of clusters $\mathcal{M}^{(3)}$, and finding their optimal values requires an exhaustive search in the 3-dimensional parameter space based on an objective function. We first describe the objective function that is based on these three parameters and then a procedure for determining the optimal values for these three parameters given a read data $\mathbf{R}$. 

We define an internal compactness measure $I-index$ as a function of two internal clustering compactness measures, Dunn Index and the Davies-Bouldin (DB) Index \cite{dunn,dbindex}. It measures the overall goodness of a clustering run where both compact and well-separated clusters are taken into account. We briefly describe the computations for Dunn and DB-index and define the ($I-index$).


The Dunn Index ($DI$) is defined as the ratio of the minimum inter-cluster distance to the maximum diameter amongst all the clusters. Here, the inter-cluster distance is defined as the distance between the consensus sequences of two clusters, and the distance between two sequences is defined as one minus the pairwise alignment scores of the two sequences. The diameter of a cluster is the average distance of all reads assigned to the cluster by $\mathit{C(\theta)}$ to the consensus sequence of the cluster. DB-Index ($DB$) is defined as an average value of the intra-cluster diameter to inter-cluster distances ratio for each cluster \cite{dbindex}. Here the diameter and the distances are defined as before, and for each cluster the maximum value of the inter-cluster diameter to inter-cluster distance is chosen in the average. The goodness of a clustering is determined by compactness of the clusters present in it which translates to small diameters, along with these clusters being well separated from each other or large value of the inter-cluster distances. Thus, high values of $DI$ index and low values for $DB$ index are indicative of a better clustering run. $DI < 1$ implies that minimum value of inter-cluster distance is smaller than the maximum diameter of any cluster, and in such a scenario a better clustering can be obtained by merging the two smaller clusters.

$I-index$ is defined as the harmonic mean of Dunn Index and inverse of DB-Index. A high value for the $I-index$ indicates a compact and well-separated set of clusters. As the measure evaluates the results obtained from a clustering run solely based on the reads, it is known as an internal clustering measure. 
\begin{equation*}
I-index = \frac{2\cdot DI}{DI\cdot DB + 1}
\end{equation*}

We use $I-index$ as an objective function for determining the optimal parameters. As the results of the clustering $\mathcal{M}^{(3)}$ are used to compute the $I-index$, it is an indirect function of the clustering parameters. The value of $T_{err}$ is fixed at 2 in the current study and we search for the optimal values of  $\{\theta^{(1)},\theta^{(3)}\}$ by performing clustering at different values. The clustering thresholds $\{\theta^{(1)},\theta^{(3)}\}$ (for step 1 and step 3 in the pipeline) which maximize the clustering compactness measure ($I-index$) and have Dunn Index ($DI$) greater than 1 ($DI > 1$) are chosen as the optimal clustering parameters for the collection of reads $\mathbf{R}$.

\vspace{-5pt}
\subsection*{External measures for cluster evaluation}
External clustering measures evaluate the clustering results based on the ground truth values known a priori. For the simulated datasets, as the number of integration sites and affiliation of reads are known, we can also use external measures to evaluate and validate the clustering results at optimal parameters. Each read in $\mathbf{R}$ is obtained from one of the $P$ integration sites. Thus, the true assignment for the reads $\mathbf{R}$ is known. 

We define a single external measure, , known as the $E-index$ (similar to V-Measure \cite{Vmeasure}), as the harmonic mean of completeness, homogeneity, and true fraction. 

\begin{equation*}
 E-index = \frac{3\cdot c\cdot h\cdot t}{c\cdot h + h\cdot t + t\cdot c}
\end{equation*}
The $E-index$, as defined above, has a value in the range of $[0,1]$, where higher values indicate better clustering. Homogeneity ($h$) for a clustering run $\mathcal{C(\theta)}$ is the fraction of the clusters that are not merged in the clustering. A cluster $M_k$ is called a merged cluster if it contains reads from two different reference integration sites. The measure $h$ indicates the fraction of clusters that contain all reads from a single reference integration site. Completeness ($c$) for a clustering run $\mathcal{C(\theta)}$ is the fraction of integration sites that are not split in the clustering. Split references have their reads clustered into two or more clusters in the clustering run. The completeness $c$, complementary to homogeneity, indicates the fraction of integration sites that have all their reads in a single cluster. The true fraction ($t$) for the clustering run $\mathcal{C(\theta)}$ is the ratio ``correct clusters'' to the total number of reference integration sites, where a ``correct cluster'' consists of a cluster $M_k$ whether all the reads from a single integration site $C_a$ and none from a different segment are present in it.

Apart from the $E-index$ for a clustering run $\mathcal{C(\theta)}$, the consensus sequence of a cluster should also be close to the original integration site sequence used to generate the reads. Thus, we also measure the percent identity of the consensus sequence of a cluster $Con(M_k)$ to the integration site sequence used to generate it. 

\vspace{-10pt}
\section{Experimental Results}
\subsection{Simulated data}
The preference of the site of insertion of mobile elements varies considerably and can include sequence specific or non-specific sites in repetitive and unique regions of the genome \cite{lewinski2005retroviral}. We therefore chose sites at random from the human genome reference (build hg19) to mimic the integration sites of any mobile element and determined the relative distribution of unique and repetitive sequences in our dataset using RepeatMasker ( \url{http://www.repeatmasker.org/cgi-bin/WEBRepeatMasker}). Two simulation data sets are generated consisting of 100 bp fragments from the human genome that represent host regions flanking the integration of the mobile element (Table \ref{Tab:01}). Sequences from 1993 locations of chromosomes 15 and 22 (dataset S1), and 2201 locations of the entire human genome (dataset S2) comprised the host flanking regions of the integration sites. We assumed for the simulation that the mobile element sequences have high identity at each site because the primer used to amplify the junction must be placed in a conserved region and close to the end of the element. We simulated Ion Torrent sequencing reads (sequencing error rate 1-2\%) from these integration sites using the simulation software dwgsim (\url{http://github.com/nh13/DWGSIM}) to an average coverage of 149 and 1318, respectively (Table \ref{Tab:01}). The reads from datasets S1 and S2 were trimmed to 40 bps for faster analysis. 


\begin{table}[!htb]
\centering
\caption{\bf{Simulated data}}
\begin{threeparttable}
\begin{tabular}{|c|c|c|c|c|} \hline
\MyHead{1.5cm}{Dataset (Source)} & \MyHead{1.75cm}{ \# of integration sites} & \MyHead{0.75cm}{\# of reads} & \MyHead{1.4cm}{\% of repeats} & \MyHead{1cm}{Avg. Coverage} \\ \hline
\MyHead{1.5cm}{S1 (chr15 \& chr22)} & 1993 & 295,985 & 23.3 & 149\\
S2(chr 1-22)& 2201 & 2,901,970 & 22.4 & 1318\\
\hline 
\end{tabular} 
\end{threeparttable}
\label{Tab:01}
\end{table}

\vspace{-5pt}  
\subsection{Clustering using the proposed pipeline}
Our goal is to establish a method that estimates optimal parameters for clustering and accurately returns the number and correct sequence of the integration sites. We applied the four-step clustering pipeline on the S1 and S2 datasets. In step one, the reads are clustered at a clustering threshold $(\theta^{(1)})$ varying from 75\% to 95\% in steps of 5 percentile. In step two, for clustering at each threshold, $(\theta^{(1)})$, we remove the clusters that contain two reads or less assigned to them ($T_{err}$ is set to two in our pipeline). In step three, we re-cluster the remaining clusters from step two at clustering threshold $(\theta^{(3)})$ varying again from 75\% to 95\% in steps of 5, so that false clusters introduced by sequencing errors are combined together (Here the superscript on the clustering threshold indicates the step of our pipeline). The consensus sequence of the clusters thus obtained in step three represent the reference integration site sequences, and we map all the reads to them in step four. Remapping of reads in step four updates the consensus sequences for each cluster and also maps the reads that were left out in step two of the pipeline. 

We evaluate the clustering performed at each clustering threshold pairs $(\theta^{(1)};\theta^{(3)})$ based on internal and external measures to determine an optimal clustering threshold. For S1, the combined internal measure, $I-index$, is maximized for clustering threshold pairs $\{ (\theta^{(1)};\theta^{(3)}) = ((80;75),(80;80))\}$, while the external clustering measure $E-index$ is maximized for clustering threshold pairs $\{ (\theta^{(1)};\theta^{(3)}) = ((75;75),(75;80))\}$. There are two sets of clustering parameters for which both the I-index and E-index are near maximum $((75;75)$ and $(80;80))$. However, the number of clusters obtained using these thresholds differs (1894 vs 1982). 

\begin{figure}[!htb]
\centering
\includegraphics[trim={25cm 0 25cm 0},scale=0.3]{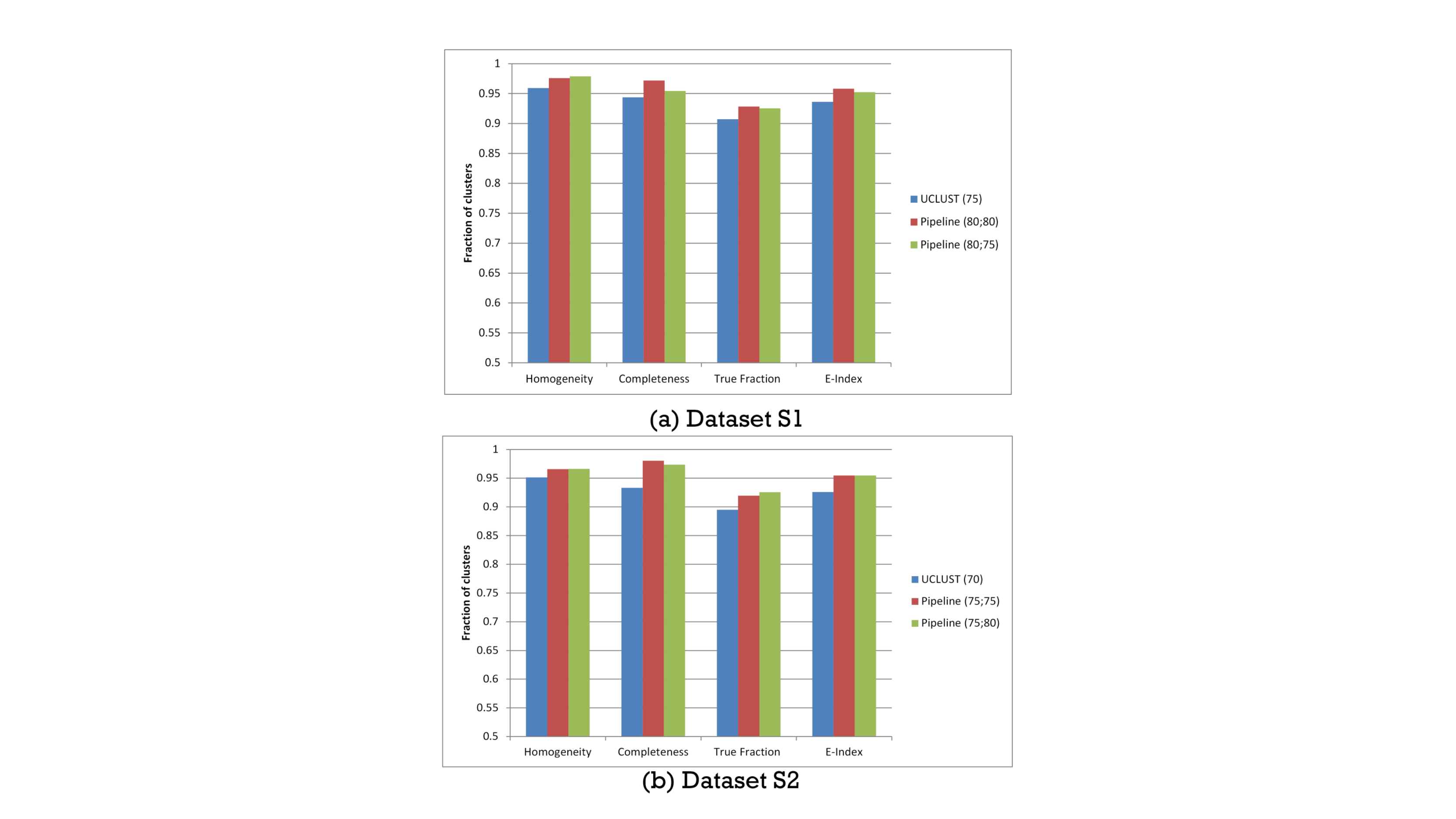}
\caption{\label{fig:05} Comparison of external clustering measures in UCLUST and the proposed pipeline for datasets S1 (a) and S2 (b). }
\end{figure}

The clustering results are also compared to those obtained at step 1, in order to assess the improvements in clustering measures from the proposed clustering pipeline.We observe that all the external clustering compactness measures, namely homogeneity, completeness, true fraction, and $E-index$, at the optimal clustering thresholds using our pipeline are better than those at optimal clustering thresholds ($\theta^{(1)} = 75$) for a single round of UCLUST clustering (Figure \ref{fig:05}). For S1 dataset, the optimal clustering parameters are $\{ (\theta^{(1)};\theta^{(3)}) = ((75;75),(75;80))\}$ based on the external clustering measures. The overall external compactness measure $E-index$ is better than that of the optimal clustering threshold from UCLUST, namely $(\theta^{(1)}=75)$, indicating overall improvement in clustering compared to a single round of UCLUST clustering.  Additionally, the number of clusters obtained after single round is 2014, which is greater than the total number of integration sites (1993) and also from the optimal results of the proposed pipeline.

\begin{figure}[!htb]
\centering
\includegraphics[width=0.4\textwidth]{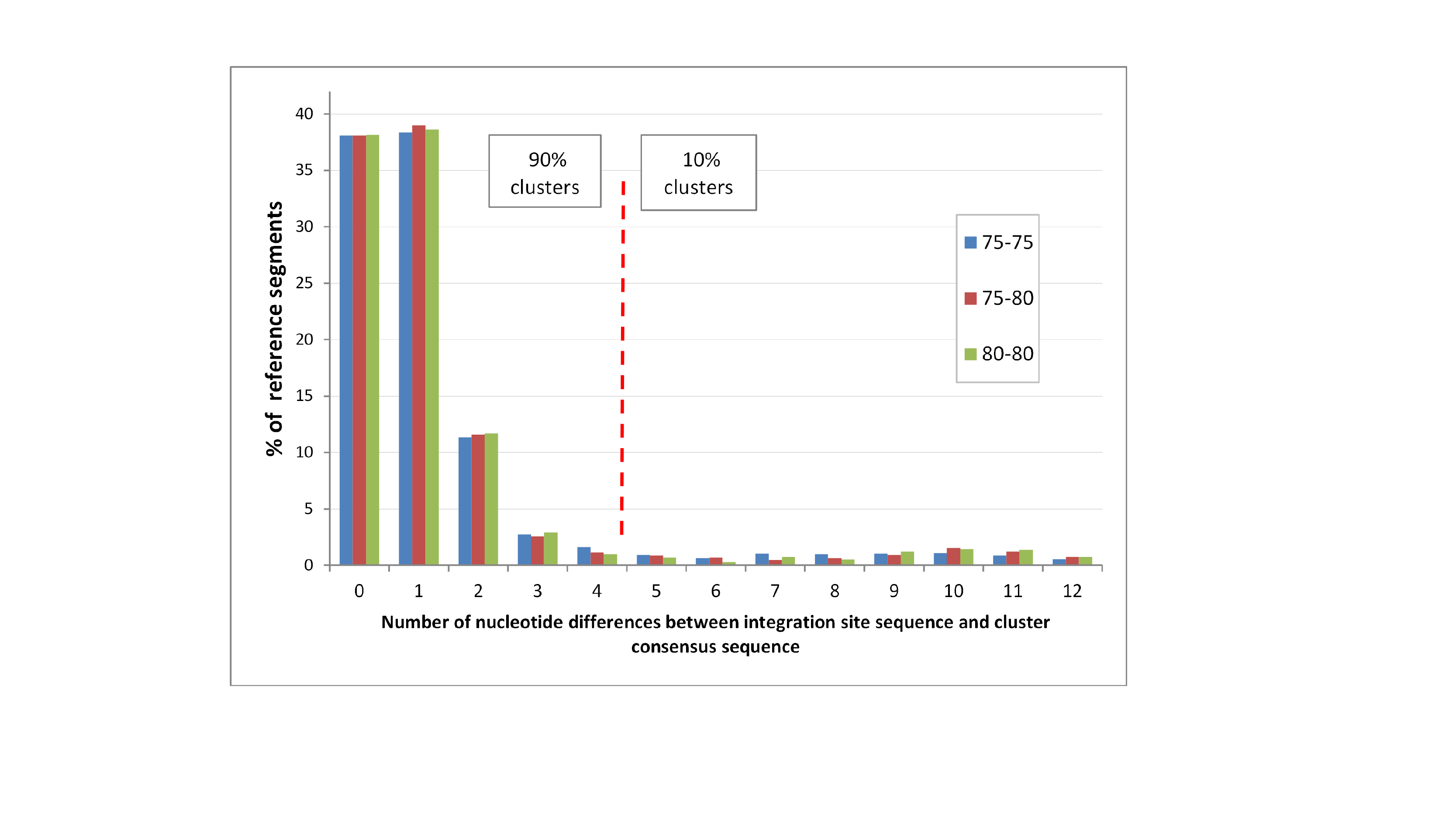}
\caption{\label{fig:04} Histogram of the number of nucleotide differences between the consensus sequence of clusters and the corresponding references for dataset S1. } 
\end{figure}

As there are multiple optimal clustering threshold pairs obtained for dataset S1, we assess the quality of clustering by comparing the consensus sequences of the clusters to the true sequences of the integration sites at these thresholds. The histogram of the base-pair sequence differences between the consensus sequences and the true sequences is similar at all optimal clustering thresholds ($\{ (\theta^{(1)};\theta^{(3)}) = ((75;75),(75;80),(80;80))\}$) in dataset S1  (Figure \ref{fig:04}), indicating that all the optimal clustering thresholds are comparable in quality, and based on a user-specified criterion, any one of them can be chosen as the clustering output. 

\textbf{Effect of Read Depth:} The dataset S2 is used to evaluate the impact of increased sequencing depth on our pipeline. Our results for clustering S2 dataset with our pipeline are similar to those obtained for S1,indicating read depth has little effect on the overall clustering accuracy. The clustering threshold pairs $\{(\theta^{(1)};\theta^{(3)})=((75;75),(70;70))\}$ maximize the combined internal measure $I-index$, which contains the clustering threshold pairs $\{(\theta^{(1)};\theta^{(3)})=(75;75)\}$ for which the external clustering index $E-index$ is also optimized. The number of clusters generated at optimal clustering threshold pairs are 94\%-98\% of the actual number of reference integration sites. 

\begin{figure}[!htb]
\centering
\includegraphics[width=0.35\textwidth]{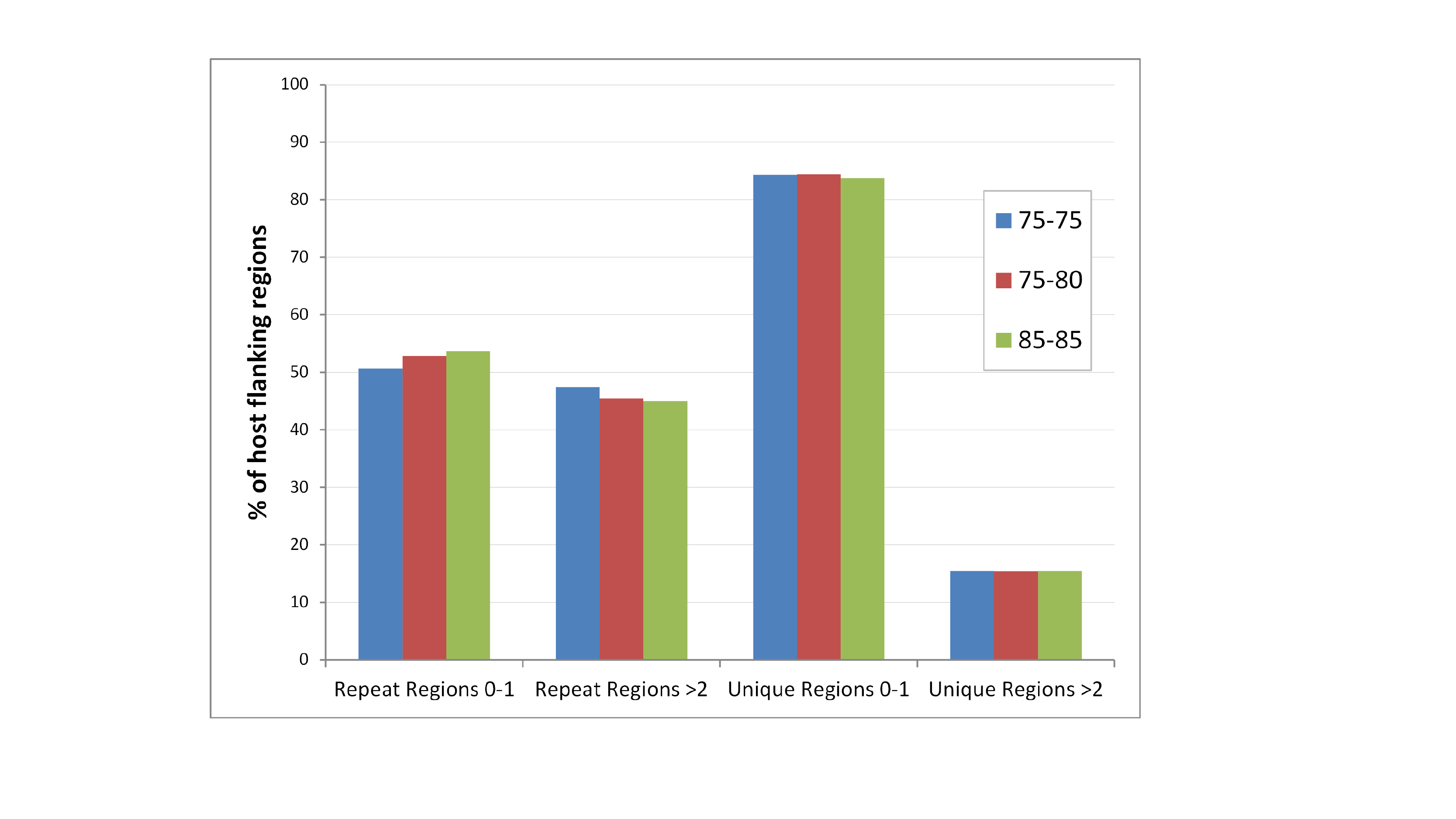}
\caption{\label{fig:06} Comparison of consensus sequences of the clusters with respect to repeat content of the references.}
\end{figure}

\textbf{Genome Content:} The simulated datasets model integration site junctions that span both the repeat regions and unique regions of the human genome. We evaluate the recovery of clusters for repeat region integration sites versus the unique region integration sites in the genome, as the repeat region integration sites have a higher likelihood of forming merged clusters. As mentioned in Table \ref{Tab:01}, 23.3\% of integration sites in S1 and 22.4\% of integration sites in S2 are from known repeat sequences. At the optimal clustering threshold pairs for S1, ($\theta^{(1)};\theta^{(3)})= (80;80)$), less than 55\% of the repeat region clusters differ by only 1 bp from their corresponding integration site (Figure \ref{fig:06}). On the other hand, 84\% of the unique region clusters are recovered to within 1 bp of their corresponding integration site. Thus the presence of integration sites spanning repeat regions of the genome have a higher impact on clustering accuracy as compared to the unique regions of the genome. The results for S2 dataset are similar to S1 data.


\vspace{-15pt}
\section{Conclusion}
We have presented a pipeline for clustering of NGS reads obtained from sequences representing the integration site of a mobile element and the flanking host region.
The pipeline proposes clusters that contains reads sampled from a single integration site with the cluster's consensus sequence denoting the sequence of the integration site-flanking host junction, using multiple rounds of clustering from a known clustering algorithm UCLUST \cite{usearch} at optimal clustering thresholds. The optimal clustering thresholds are empirically determined by maximizing our proposed $I-index$ over a combination of clustering thresholds. Our results demonstrate that these parameters are dependent on the sequence content of the integration sites. Compared to a single round of UCLUST clustering, the total number of clusters proposed by our pipeline approximates the actual number of integration site sequences in the simulated datasets, while showing improvements in the clustering measures $I-index$ and $E-index$. The proposed pipeline is insensitive to the sequence depth of the dataset and recovers compact and well-separated clusters in our two datasets which differ by nine-folds in sequence coverage. Thus, a small fraction of a real dataset can be used to quickly determine optimal clustering parameters, as the clustering pipeline is dependent on the number of sequences.  The pipeline has applications for identifying sequences from integration site junctions, where the primer representing the mobile element is used to sequence a part of the mobile element and the flanking host region. Thus, as the sequence of the mobile element is relatively conserved, all the reads start with the same sequence, and the reads originating from a single host-mobile element junction can be clustered together efficiently if the reads are trimmed to a constant length. The constant length of reads helps to mitigate the insertion or deletion errors one observes in the PCR sequencing. Ensuring that the reads are pre-processed to remove the primer sequences and that they start with a consistent sequence of the mobile element is beneficial for clustering, although not necessary. Our approach provides a quantitative clustering method that has downstream applications in studying the distribution of polymorphic insertions of mobile elements among individuals, where it is essential to determine the exact number of and the sequence of the insertion site junctions.

\vspace{-10pt}
\section*{Acknowledgments}
This work was supported, in part, by the National Science Foundation Awards 1421908, 1533797, the United States Geological Survey award 06HQAG0131 and by the Czech Ministry of Education, Youth and Sports grant LK11215.

\bibliographystyle{plain}

\end{document}